\documentstyle[epsfig,12pt]{article}

\newcommand{\nc}{\newcommand}
\nc{\beq}{\begin{equation}}   \nc{\eeq}{\end{equation}}
\nc{\beqa}{\begin{eqnarray}}  \nc{\eeqa}{\end{eqnarray}}
\nc{\lsim}{\begin{array}{c}\,\sim\vspace{-21pt}\\< \end{array}}
\nc{\gsim}{\begin{array}{c}\sim\vspace{-21pt}\\> \end{array}}
\nc{\create}{\hat{a}^\dagger}   \nc{\destroy}{\hat{a}}
\nc{\kvec}{\vec{k}}             \nc{\kvecp}{\vec{k}^\prime}
\nc{\kvecpp}{\vec{k}^{\prime\prime} }   \nc{\kb}{\bf k}
\nc{\kbp}{{\bf k}^\prime}       \nc{\kbpp}{{\bf k}^{\prime\prime}
}
\nc{\bfk}{{\bf k}}              \nc{\cohak}{a_{{\bf k}}}
\nc{\cohap}{a_{{\bf p}}}        \nc{\cohaq}{a_{{\bf q}}}
\nc{\cohbk}{b^*_{{\bf k}}}      \nc{\cohbp}{b^*_{{\bf p}}}
\nc{\cohbq}{b^*_{{\bf q}}}

\begin{document}

\title{{\large {\bf Quantum Production of Black Holes}}}
\author{
Stephen D.H.~Hsu\thanks{hsu@duende.uoregon.edu} \\
Department of Physics \\
University of Oregon, Eugene OR 97403-5203 \\}
\date{March, 2002}
\maketitle

\begin{abstract}
We give a path integral expression for the quantum amplitude to
produce a black hole from particle collisions. When expanded about
an appropriate classical solution it yields the leading order
contribution to the production amplitude in a curvature expansion.
Classical solutions describing black hole production resulting
from two particle scattering at non-zero impact parameter,
combined with our formalism, indicate a geometric cross section
for the quantum process. In TeV gravity scenarios these solutions
may exhibit large curvatures, but (modulo a mild assumption about
quantum gravity) corrections to the semi-classical cross section
are small.
\end{abstract}

\newpage

\section{Introduction}

Recently proposed models with extra dimensions solve the hierarchy
problem by bringing the fundamental scale of gravity (henceforth
referred to as the Planck scale) down to the electroweak scale
\cite{tevgrav}. In such scenarios, quantum gravitational effects
arise at energies as low as a TeV. Perhaps the most dramatic
example of such phenomena is black hole production in particle
collisions with center of mass energy greater than the Planck
scale \cite{ACV},\cite{TH}. Such events would lead to dramatic
signatures at colliders and in cosmic ray collisions, and perhaps
imply an end to our ability to probe shorter and shorter length
scales \cite{BF}-\cite{phenom}.

The cross section for black hole production in high energy
collisions is difficult to compute. In \cite{ACV}-\cite{phenom} it
was asserted that the cross section is geometrical, determined by
the impact parameter at which the particle pair at closest
approach is within the Schwarschild radius associated with the
center of mass energy $\sqrt{s}$. If this is the case, black holes
would be copiously produced at LHC and in cosmic ray collisions.
However, Voloshin has criticized these claims, arguing for an
exponentially small cross section \cite{voloshin}.

Eardley and Giddings \cite{eg} have analyzed classical solutions
in general relativity which describe two particle high energy
collisions at non-zero impact parameter (see also
\cite{Aichelburg:1970dh}-\cite{CL}). They demonstrate the
existence of a closed trapped surface for any collision with
sufficiently small impact parameter (at fixed center of mass
energy). Their lower bound on the critical impact parameter leads
to a geometrical classical cross section in rough agreement with
the earlier naive estimates.

The solutions in \cite{eg} yield immediate answers to Voloshin's
two main objections: (see also \cite{SS} for a model calculation
that shows no exponential suppression)

1) Euclidean suppression: because there are classical trajectories
with two particle initial conditions which evolve into black
holes, the process is clearly not classically forbidden, and hence
there is no tunnelling factor.

2) CPT (time reversal): Voloshin argues that since black holes
produce a thermal spectrum of particles during evaporation, rather
than a few highly energetic particles, the time-reversal of the
production process (and hence the production process itself, by
CPT) must have very low probability. However, the time-reversed
classical solutions exhibit a very energetic wave of gravitational
radiation colliding with the time-reversed black hole to produce
the two particle state (in the formation process this is the
energy which escapes the hole). The process is not thermal and
involves very special initial (final) conditions.

In this letter we use a path integral formalism previously
developed by Gould, Hsu and Poppitz \cite{ghp,nato} (GHP) to study
non-perturbative scattering. In the GHP formalism quantum S-matrix
elements are computed in a systematic expansion about classical
solutions satisfying appropriate boundary conditions. We show that
the formalism is readily adapted to the problem of black hole
production from high energy collisions.

Although some black hole production amplitudes (for example,
involving many soft initial particles) can be computed
unambiguously from classical solutions, in the two particle case
quantum corrections might be significant due to large curvatures.
However, we argue that even in this case the semi-classical
approximation is probably a good one. We explain the importance of
large curvatures to the quantum corrections from an effective
field theory point of view.

\section{Path Integral Formalism}

In this section we review a method formulated to describe
scattering processes involving non-perturbative field
configurations using the path integral \cite{ghp,nato}. Previous
interest in this problem centered on baryon number violation in
the electroweak theory. In this context, it was realized that
classical configurations satisfying vacuum boundary conditions
(such as instantons) were unsuitable for the computation of high
energy scattering \cite{MMY}. A good semiclassical approximation
in this context requires taking into account the initial state
overlap as well as the classical action.

In \cite{ghp}, an {\it exact} expression is derived for the
S-matrix describing a transition from an initial two particle
state to a coherent final state. (See below for a precise
definition of coherent state.) This expression is approximated in
the usual stationary phase approximation, which leads to a
boundary value problem involving the usual classical equations of
motion, but with boundary conditions determined by the initial and
final states. Unfortunately, numerical searches for
``interesting'' solutions of this boundary value problem in
lattice gauge theories have only uncovered configurations
describing transitions between states consisting of many soft
quanta \cite{Duke}. The Eardley-Giddings result is the first we
are aware of in which a high momentum initial state evolves into a
large, low momentum final state. However, we should note that the
two particles in the initial state are ``dressed'' by strong
gravitational fields, so the effective number of gravitons in the
initial state is actually large.

We first give a brief review of the general method, before
addressing issues particular to general relativity and black
holes. In the path integral representation of the process $\vert i
\rangle ~\rightarrow~ \vert f \rangle$, trajectories are weighted
by the appropriate action $\exp [ i S ]$, as well as by the
overlap of the asymptotic part of the path with the initial and
final states. Therefore we expect that the amplitude must be
expressible in the form \cite{nato}: \beq \label{heur} \langle f
\vert S \vert i \rangle ~\sim~ \int d \phi_i~ d \phi_f ~D \phi~
\Psi_i [ \phi(T_i)] ~ \Psi_f^* [\phi(T_f)]~ e^{i S[\phi]}, \eeq
where we have explicitly indicated the fluctuations of the fields
in the asymptotic past and future ($T_{i,f}$) in the measure. The
wave-functionals $\Psi_{i,f}$ measure the overlap of the initial
and final states with $\hat{\phi}$ eigenstates at asymptotic
times. In \cite{ghp} a derivation of (\ref{heur}) was given, along
with the explicit form of the wavefunctionals in the case that the
initial and final states are either wave-packets (including plane
waves) or coherent states.

In the limit $T_{i,f} \rightarrow \pm \infty$, the amplitude in
(\ref{heur}) is just the S-matrix, $S_{fi}$. The GHP procedure
amounts to a semi-classical evaluation of this object taking into
account the initial and final state overlaps in the extremization.
The result is a boundary value problem with boundary conditions
determined by the initial and final state, but governed by the
usual equations of motion. Here our goal is to calculate the
\mbox{S-matrix} element between an initial two particle state and
a final state which includes a black hole.

First, we express the kernel of the \mbox{S-matrix} in a basis of
coherent states (see, e.g., the text by Fadeev and Slavnov
\cite{FStext}). The initial and final states are defined by sets
of complex variables ${\bf a}\equiv \{\cohak\}$, ${\bf b}^*\equiv
\{\cohbk\}$, respectively. A coherent state $\vert \cohak \rangle$
is an eigenstate of the annihilation operator $\destroy_{\bf k}$:
$\destroy_{\bf k} \vert \cohak \rangle = \cohak  \vert \cohak
\rangle$. Recall that coherent states saturate minimum-uncertainty
bounds, and hence are good (although not unique) candidates for
semi-classical states.

The transition amplitude from an initial coherent state $\vert
{\bf a}\rangle$ at time $T_i $ to a final coherent state $\vert
{\bf b}^*\rangle $ at time $T_f $, can be expressed \beq
\label{transamp} \langle\, {\bf b}^*\,\vert \, U\,\vert\, {\bf
a}\,\rangle ~=~ \int d\phi_i\, d\phi_f ~\langle\, {\bf
b}^*\,\vert\,\phi_f\,\rangle~ \langle\,\phi_f\,\vert\,
U\,\vert\,\phi_i\,\rangle~ \langle\,\phi_i\,\vert\, {\bf
a}\,\rangle ~, \eeq where $U$ is the evolution operator between
time $T_i $ and $T_f $. A``position'' eigenstate of the field
operator $\phi $ is denoted $\vert\,\phi\,\rangle $ and
$\phi_{i,f}=\phi(T_{i,f}) $. Then, from (\ref{transamp}), we
obtain the \mbox{S-matrix} kernel in a compact form in terms of
path integrals \beq \label{s} \langle\, {\bf b}^*\,\vert\,
S\,\vert\, {\bf a}\,\rangle ~\equiv~ S\left[ {\bf b}^*, {\bf a}
\right] ~=~ \lim_{T_i,T_f \rightarrow \mp \infty}~ \int d\phi_f\,
d\phi_i~ e^{B_f}~e^{B_i} \int_{\phi_i}^{\phi_f} D\phi~
e^{iS\,\left[\,\phi\,\right]} ~, \eeq where $S\left[\phi\right]$
is the action functional. The path integral appearing here is over
fields obeying the boundary conditions $\phi(T_{i,f}) =
\phi_{i,f}$. The functional $B_f$ is \beqa \label{bf} B_f\left[
{\bf b}^*, \phi_f\right] & = & - {1\over 2} \int\! d^3k ~ \cohbk
b^*_{\bf -k} ~e^{2i\omega_kT_f} - {1\over 2} \int\! d^3k
~\omega_{\bf k} ~\phi_f( \kvec )
{}~\phi_f(-\kvec ) \\
&  & +~\int\! d^3k ~\sqrt{2\omega_{\bf k}}~e^{i\omega_k T_f}~
\cohbk~\phi_f(-\kvec ) \nonumber ~, \eeqa in terms of which the
wave functional of the final coherent state is $\langle\, {\bf
b}^*\,\vert\,\phi_f\, \rangle ~\equiv~ \exp{\left(\, B_f\left[
{\bf b}^*,\phi_f\right] \,\right)} ~.$ Similarly, the functional
$B_i$ can be expressed in terms of the initial coherent state
wavefunctional $\langle\,\phi_i\,\vert\, {\bf a}\,\rangle ~\equiv~
\exp{\left(\, B_i\left[ {\bf a},\phi_i\right] \,\right)}$, with
$b^*_{\bf k}$ replaced by $a_{\bf k}$ and $T_f$ by $- T_i$ in
(\ref{bf}). The 3-dimensional Fourier transform is defined \beq
\label{FT} \phi_{i,f}( \kvec ) ~=~ \int {d^3x \over (2\pi)^{3/2}
}~ e^{i\vec{k}\cdot\vec{x}} ~\phi (T_{i,f}, \vec{x}) ~. \eeq

The kernel (\ref{s}) is a generating functional for
\mbox{S-matrix} elements between any initial and final $N$
particle states, by functional differentiation with respect to
arbitrary $\cohak$ and $\cohbk$. We now use this fact to construct
a kernel for scattering from initial two particle states. We
define an initial two particle (wave packet) state at $t = T_i$
\beq \label{alpha} \vert\,\vec{p}, -\vec{p}\,\rangle ~\equiv~ \int
d^3k ~\alpha_R(\kvec ) ~\create_{\bf k} \int d^3k^\prime
~\alpha_L(\kvecp)~\create_{\kbp}~ \vert\, 0\, \rangle ~, \eeq
where $\create_{\kb} $ is a creation operator, and
$\alpha_{R,L}(\kvec )$ are arbitrary smearing functions of $\kvec
$, localized around some reference momenta $\vec{p}$ and
$-\vec{p}$ respectively. The wave packets are normalized so that
$\int d^3k ~\vert\, \alpha_{R,L}(k )\, \vert^2 ~=~ 1 ~$.

This state can be generated by functional differentiation of the
coherent state $\vert {\bf a}\rangle $ with respect to $ a_{\bf k}
$ \beq \label{f2} \vert\vec{p}, -\vec{p}\rangle ~=~ \int
d^3\kvec~d^3\kvecp~ \alpha_R(\kvec ) ~\alpha_L(\kvecp)~
{\delta\over\delta a_{\bf k}}~{\delta\over\delta a_{\kbp}}~
\vert\, {\bf a}\,\rangle ~\rule[-3mm]{0.2mm}{9mm}_{~{\bf a}\, =\,
0} {}~. \eeq

So, differentiating under the functional integral, the
\mbox{S-matrix} element between the two particle state (\ref{f2})
and any final state $\vert\{\cohbk\}\rangle$ involves the
following functional at $t = T_i$ \beqa \label{bi2} \lefteqn{
{\delta\over\delta a_{\bf k}} {\delta\over\delta a_{\kbp}}
\exp{\left(\, B_i\left[{\bf a},\phi_i\right] \,\right)}~
\rule[-3mm]{0.2mm}{9mm}_{~a = 0} ~ =} & & \\
& & \hspace{2cm} 2~\sqrt{\omega_{\bf k}\omega_{\kbp}}~
\phi_i(\kvec)~\phi_i(\kvecp)~e^{-i(\omega_{\bf
k}+\omega_{\kbp})T_i} \exp{\left(\, B_i\left[0, \phi_i\right]
\,\right)}\nonumber ~, \eeqa after dropping a term which vanishes
in the limit $T_i \rightarrow -\infty$. The last factor here is
simply the normalization of the initial position eigenstate \beq
\exp{\left(\, -{1\over 2}\int d^3k~\omega_k~\phi_i(\vec{k}
)~\phi_i(-\vec{k} ) \,\right)} ~. \eeq

We combine this with the smearing functions and finally obtain an
\mbox{S-matrix} kernel for the scattering of two wave packets into
arbitrary final states, \beq \label{combine} S\left[{\bf b}^*, 2
\right] ~=~ \lim_{T_i,T_f\rightarrow \mp\infty} \int d\phi_f\,
d\phi_i~ \alpha_R\cdot\phi_i~\alpha_L\cdot\phi_i ~ e^{B_f\left[
b,\phi_f\right] ~+~ B_i\left[0, \phi_i\right] }
\int_{\phi_i}^{\phi_f} D\phi~e^{iS\,\left[\,\phi\,\right]}~, \eeq
where we have denoted the initial state (\ref{f2}) by ``$2$''.
Here we have used the following compact notation for the initial
state factors \beq \label{integral} \alpha\cdot\phi_i ~\equiv~
\int d^3k ~\sqrt{2\omega_k}~ \alpha(k ) ~\phi_i(k ) ~e^{-i\omega_k
T_i} ~. \eeq

We exponentiate the initial state factors into an ``effective
action'', so that \beq \label{seff} S\left[ {\bf b}^*, 2\right]
~=~ \lim_{T_i,T_f\rightarrow\mp\infty} \int d\phi_f\, d\phi_i\,
D\phi ~e^{\Gamma}  ~, \eeq where the effective action $\Gamma$ is
\beq \label{gam} \Gamma\,\left[\,\phi\,\right] ~=~ \ln
\alpha_R\cdot\phi_i~\alpha_L\cdot\phi_i ~+~ B_i\left[0,
\phi_i\right]~+~ iS\,\left[\,\phi\,\right] ~+~ B_f\left[b^* ,
\phi_f\right] ~, \eeq after dropping a term which vanishes as
$T_i\rightarrow -\infty$.

We can now derive the boundary value problem by varying the
effective action. Varying the entire exponent $\Gamma$ with
respect to $\phi(x)$ for $T_i < t < T_f$ gives the  source-free
equations of motion \beq \label{eom} {\delta S \over \delta
\phi(x)} ~=~ 0 ~. \eeq Varying the entire exponent with respect to
$\phi_i(k)$, gives
\beq \label{ti} i\,\dot{\phi}_i (\vec{k}) ~+~
\omega_k \phi_i (\vec{k}) ~=~ \sqrt{2\,\omega_{\bf k}}~\left(\,
{\alpha_R(\vec{k})\over\alpha_R\cdot\phi_i}~+~
{\alpha_L(\vec{k})\over\alpha_L\cdot\phi_i}\,\right)
{}~e^{-i\omega_k T_i} ~. \eeq
The first term on the left hand side
comes from a surface term in the action $S$. The other terms come
from variation of the wave functional at $t=T_i$. This boundary
condition involves both the positive and negative frequency parts
of the field.

The boundary condition (\ref{ti}) at the initial time slice is
rather complicated. However, it can be simplified since a real
field $\phi$ may be written in the asymptotic region $t =
T_i\rightarrow -\infty$ as a plane wave superposition \beq
\label{phialpha} \phi_i(\vec{k}) ~=~ {1\over \sqrt{2\,\omega_{\bf
k}}}~ \left(\, u_{\bf k}~e^{-i\omega_k T_i} ~+~ u_{-\bf
k}^*~e^{i\omega_k T_i} \,\right) ~. \eeq Equation (\ref{ti}) then
reduces to the requirement \beq \label{u} u_{\bf k} ~=~ {
\alpha_R(\vec{k} ) ~+~ \alpha_L(\vec{k}) \over \left(\, 1 ~+~ \int
d^3k ~\alpha_R(\vec{k} )~\alpha_L(\vec{k}) \,\right)^{1/2} }~,
\eeq using the normalization condition on $\alpha_{L,R}$. This
solution is consistent with physical intuition, the classical
field reducing to the initial particles at early times. The
overlap of the left- and right- moving wave packets in the
denominator is very small for narrow high energy wave packets.

A similar analysis relates the late time boundary condition on
$\phi$ to the coherent state ${\bf b}^*$ \cite{ghp}. The classical
field satisfying these boundary conditions extremizes the S matrix
for production of the coherent state in a two particle collision:
$S [ {\bf b}^*, 2]$. Note that for arbitrary choice of ${\bf b}^*$
there is no guarantee of a classical solution satisfying the
necessary boundary conditions: in some cases a complex trajectory
extremizes the S-matrix, leading to exponential suppression of the
process \cite{ghp}. However, conversely {\it every} classical
solution obeying initial conditions (\ref{phialpha})-(\ref{u})
corresponds to an {\it unsuppressed} quantum amplitude.

Now consider the extension of this formalism to general
relativity, and to black hole production. Clearly one can replace
$\phi$ with the metric field $g_{\mu \nu}$ plus appropriate matter
fields. While general covariance does not permit a unique time
slicing, the gravitational action $S_g = \int d^4x \sqrt{-g}~ R~$
is still well-defined and, in fact, the action can be expressed in
terms of a surface integral over Bondi masses \cite{surface}.

The notion of an S-matrix is appropriate if we consider
asymptotically flat spacetimes in the far past and future, plus
additional excitations. Black holes themselves are considered
excitations, and we must extend our Hilbert space to include
quantum states representing black holes. A pragmatic way to
approach this is to define (semi-classical) black hole states as
those with a strong overlap with the trajectories corresponding to
classical black holes. (These must of course have mass much larger
than the Planck mass.) In a classical black hole solution, excess
energy is radiated away by late times and the exterior metric can
be classified by a limited number of quantum numbers such as mass,
charge and angular momentum. A minimal formulation involves a
Hilbert space of black holes classified by their exterior metric
at future null infinity.

Although it is beyond the scope of this letter, it is worth noting
that a more detailed analysis of black hole states, which takes
account of the {\it internal} structure of the black hole (i.e.
the fields inside the horizon, not just at future null infinity),
indicates more states than are counted by externally visible
quantum numbers. That is, if one considers the internal structure
of the black hole in defining the Hilbert space, there are many
additional semi-classical states that one might associate with a
given exterior metric, but which differ radically within. Gedanken
experiments involving semi-classical black holes in this formalism
might teach us something about black hole information. For
example, relative phases and interference patterns due to internal
structure might be observable in black hole scattering.

Any classical solution connecting particle-like initial conditions
to an asymptotic black hole configuration provides an extremal
configuration about which to expand the S-matrix; the leading
contribution is a pure phase with no exponential suppression
\cite{ghp}. In the next section we consider quantum corrections to
this leading semi-classical approximation.

\section{Quantum corrections}

In the original application of \cite{ghp} to quantum fields in
flat spacetime, it was shown that the quantum corrections to the
semi-classical approximation to the S-matrix are suppressed by
powers of the coupling constant. In particular, the corrections
can be expressed in terms of propagators and interaction terms
which result from expanding about the classical solution. All
interaction terms carry explicit powers of the coupling, resulting
in a well-defined loop expansion.

In general relativity there is of course no small dimensionless
parameter. Expanding about a background configuration yields
interactions which are suppressed by the background curvature in
Planck units. Classical solutions describing the ordinary
gravitational collapse of many ``soft" particles (e.g., collapse
of a large star or dust ball) can produce black holes without
regions of large curvature. Our formalism applies directly to such
solutions, resulting in a semi-classical amplitude without large
quantum corrections.

In the two-particle solutions of \cite{eg}, regions of large
curvature can arise quite early in the evolution (e.g., when shock
fronts collide), even if the black hole produced is large
($\sqrt{s} >> M_{\rm Planck}$). If one takes the size of the
colliding particles to be of order the Planck length L, one finds
curvatures at the shock front of order $s$. In this case, quantum
corrections might be large. In fact, we run into a fundamental
problem concerning quantum gravity. Because gravity is
non-renormalizable, we have to consider all possible generally
covariant higher dimension operators in our lagrangian, such as
higher powers of the curvature. Certainly, such terms will arise
from the ultraviolet part of any loop calculation and will
presumably be only partially cancelled by counterterms. In an
effective lagrangian description, we expect these operators to be
present, but suppressed by powers of the Planck scale. In large
curvature backgrounds these terms may not be negligible, so the
size of quantum corrections will in principle depend on unknown
details of quantum gravity.

We can state this conclusion in a slightly different way. Consider
the {\it classical} evolution of some initial data in an effective
low-energy description of gravity. We can only trust the Einstein
equations (which result from the lowest dimension term in the
effective lagrangian) if large curvatures are {\it never}
encountered during the evolution. Once a region of large curvature
is encountered, subsequent evolution might depend in detail on the
nature of the higher dimension operators, and hence on the nature
of quantum gravity\footnote{This is quite similar to the case of
long-wavelength classical configurations in the QCD chiral
lagrangian \cite{dcc}. There, one must be sure that no
high-frequency bunching of modes occurs; otherwise the evolution
becomes sensitive to higher order terms.}. In \cite{eg} it is
suggested that since a closed trapped surface is identifiable in a
low-curvature region, the classical solution is a good guide to
the true quantum behavior. Strictly speaking, this is not
sufficient -- the solution has already evolved through a
potentially high curvature shockwave region.

It is reasonable to expect that short distance features of the
metric which lead to high curvatures do not affect the classical
Einstein evolution of the solutions in \cite{eg} on large length
scales, such as of order b, the impact parameter. The large
distance behavior of the Aichelburg-Sexl metric
\cite{Aichelburg:1970dh} used in \cite{eg} is independent of the
short distance features of the particle as long as its size r is
much smaller than b. In fact, Kohlprath and Veneziano have
recently extended the Eardley-Giddings construction to the case of
colliding particles of finite size \cite{KV}.

In TeV gravity scenarios r is likely to be the Planck length.
However, we can instead consider collisions of particles of size
much larger than the Planck length but much less than b. (In other
words, ``colliding Jupiters'' at ultra-relativistic velocities,
with b much larger than the radius of Jupiter!) By adjusting the
impact parameter and particle size relative to the Planck length
(while keeping a large hierarchy between the three) we can keep
the shockwave curvatures parametrically smaller than $M_{\rm
Planck}^2$ while preserving the long distance behavior that leads
to horizon formation. In Planck units, the maximum curvature in
the shockwave is $R \sim (b/r)^2 (L/r)^2$, where $b \sim L^2
\sqrt{s}$ and L is the Planck length. By independently varying
$\sqrt{s}$ and r, we can make the first ratio large and the second
small, while keeping their product small. The semi-classical
approximation applies quite well in this limit and higher
dimension operators can be neglected.

Barring unexpected quantum gravitational effects which are
sensitive to the size of the objects colliding (i.e. that make
long distance behavior dependent on the short distance metric),
the quantum cross section for black hole production will be well
approximated by the semi-classical one even in TeV gravity
scenarios.

\vskip 1 in

\centerline{\bf Acknowledgments}

The author thanks Dave Soper for stimulating his interest in this
problem, and Doug Eardley and Steve Giddings for discussions. This
work was supported in part under DOE contract DE-FG06-85ER40224.


\end{document}